\documentstyle[aps,prl,epsf]{revtex}
\begin{document}
\draft
\wideabs{
\title{Anomalous magnetic splitting of the Kondo resonance}
\author{Joel E. Moore and Xiao-Gang Wen}
\address{Department of Physics, Massachusetts Institute of Technology,
Cambridge, MA 02139}
\maketitle
\begin{abstract}
The splitting of the Kondo resonance in the density of states of an
Anderson impurity in finite magnetic field is calculated from the
exact Bethe-ansatz solution.  The result gives an estimate of the
electron spectral function for nonzero magnetic field and
Kondo temperature, with consequences for transport experiments on
quantum dots in the Kondo regime.  The strong correlations
of the Kondo ground state cause a significant low-temperature reduction of the
peak splitting.  Explicit formulae are found for
the shift and broadening of the Kondo peaks.
A likely cause of the problems of large-$N$ approaches
to spin-$\frac{1}{2}$ impurities at finite magnetic field is suggested.
\end{abstract}
\pacs{PACS numbers: 72.15.Qm, 73.20.Dx, 73.40.Gk, 73.50.Fq
}}
The physics of a magnetic impurity in a metallic host has been studied
for many years, and the nature of the ground state and lowest excited states
are now understood through the numerical renormalization group~\cite{wilson}
(NRG)
and the Fermi-liquid theory of the Kondo fixed point.~\cite{noz0,noz}
The theoretical situation is much less clear for
excitations away from the Fermi surface; however, it is precisely these
excitations which are probed by two kinds of current experiments.  The
nonequilibrium conductance through a quantum dot in a magnetic field $H$,
believed to be the clearest sign of Kondo physics in such systems,
is determined by the electron spectral function on the dot away from the
Fermi energy.
The Kondo peak in the spectral function at the Fermi
level splits for $H>0$ into two peaks which move away
from the Fermi level with increasing $H$. 
The subject of this paper is an approximate calculation
of this spectral function, based on the exact
Bethe-ansatz solution~\cite{andrei,wiegmann} of the
$s$-$d$ model.


The differential conductance through a quantum dot at finite bias but
$H=0$ was calculated by Meir,
Wingreen, and Lee~\cite{mwl} using the non-crossing approximation
(NCA)~\cite{bickers}
applied to the Anderson Hamiltonian
\begin{eqnarray}
H &=& \sum_{\sigma,k}\epsilon_{k,\sigma}c^\dagger_{k,\sigma}c_{k,\sigma}
+\sum_{\sigma}\epsilon_d c^\dagger_{d,\sigma} c_{d,\sigma} + \cr
&\hphantom{=}&U n_{d,\uparrow} n_{d,\downarrow} +
\sum_{\sigma,k} (V_{k,\sigma} c^\dagger_{k,\sigma} c_{d,\sigma} + \rm{h.c.})
\label{anderson}
\end{eqnarray}
The NCA is a large-$N$ technique which sums a class of diagrams
sufficiently general to resolve the Kondo temperature $T_0$.  However,
at finite magnetic field the NCA was found to give spurious peaks
in the density of states, and thus a perturbative equations-of-motion (EOM)
approach was used in which the energy scale $T_0$ does not exist (i.e.,
is effectively zero).
The resulting splitting of the Kondo-like peaks is exactly twice the
$s=\frac{1}{2}$ Zeeman
splitting.
One motivation for this work was to determine whether the strong
correlations of the Kondo ground state cause
corrections to the peak splitting.

We develop a nonperturbative estimate of
the Kondo peak splitting and shape for all values of the ratio
$g \mu H/T_0$.  The decreased splitting can be pictured as resulting from
the oscillation of the impurity spin due to its interaction with the lead
electrons (the time scale for this oscillation is ${T_0}^{-1}$ for $H=0$).

We use the Bethe-ansatz solution to
calculate exactly the density of states for ``spinon'' excitations,
which are the only excitations at energies up to $g \mu H$ from
the Fermi level.~\cite{tsvelick}
The density of states for spinon excitations gives
an estimate for the electron impurity spectral function.
This approximate technique is
known at $H = 0$ to give results consistent with Fermi-liquid perturbation
theory and the NRG.~\cite{kawakami}
Analytic expressions are derived for zero temperature, and we comment
on the expected changes when $T>0$; the results on spinon excitations
given below can in principle be combined with existing
numerical methods for the finite-temperature thermodynamic Bethe
ansatz.~\cite{rajan}
There are simple limiting forms for the peak location and broadening
when $g \mu H \gg T_0$.  In this limit the EOM approach of \cite{mwl}
slowly becomes valid as the Kondo energy scale disappears.

The Bethe-ansatz solutions of the $s$-$d$ and Anderson models give exact
results for thermodynamic properties such as magnetization and specific
heat.  Dynamical quantities such as the electron Green's
function on the impurity site in the Anderson model
are not known exactly from the Bethe ansatz
(or from any other technique) but are required to understand tunneling and
other experiments.  For example, the Landauer-type formula for nonequilibrium
conductance~\cite{meir} depends on the spectral function
$\rho_{d,\sigma}(\omega) = (-1 / \pi)
{\rm Im} {\tilde G}^R_{d,\sigma}(\omega)$ over all energies
between the source and drain voltages.

The spectral function can be written as a sum over energy eigenstates
\begin{eqnarray}
\rho_{d,\sigma}(\omega) = \sum_n \left[
|\langle 0 | c_{d,\sigma} | n \rangle|^2 \delta(\omega - (E_n - E_0)) +
\right. \nonumber \\
\left. |\langle 0 | c^\dagger_{d,\sigma} | n \rangle|^2 \delta(\omega +
(E_n-E_0)) \right]
\end{eqnarray}
At zero magnetic field, an accurate description of the
shape of the Kondo peak in $\rho(\omega)$ follows from
assuming the matrix elements to be some constant $C$ for all states connected
to the ground state by a single
spinon (defined below), and 0 for other states.
This is successful for two reasons:
because the Kondo peak in the spectral function arises from the Kondo
peak in the DOS rather than from structure in the matrix elements,
and because spinons are the excitations onto which the electron
operator projects most strongly.  At $H>0$ there is the
additional advantage that all excitations other than spinons acquire
a gap $g \mu H$ and do not affect the spectral function below this energy.
The value of the constant $C$, which determines the overall weight in the
Kondo peak, can be found for the symmetric model ($\epsilon_d = -U/2$)
from the value of the spectral function
at the Fermi level $\rho_{d,\sigma}(0) = 1 / \pi \Delta$,
with $\Delta = \pi \rho_0 \sum |V_k|^2$ the bare linewidth.~\cite{langreth}


The DOS in the exact solution of the Anderson model
consists of a Kondo peak
near zero energy from spinons, and a peak at energy $U$ from ``holons''
(the simplest excitations in the charge sector).  In the Kondo regime,
the Anderson model reduces to the $s$-$d$ model (an impurity spin with
interaction term $J {\bf s}_i \psi^\dagger {\bf \sigma}_i \psi$,
$\psi$ the conduction electron field) if we are concerned with
energies much less than $U$ above the ground state.  Hence, to understand
the splitting of the Kondo peak for $0<g\mu H\ll U$, the spin excitations are
the relevant ones and the $s$-$d$ model is appropriate.  Mixed-valence
behavior requiring a full Anderson model description is
expected to begin at $g \mu H \geq (U \Delta)^{1/2}$, which is a very high
field for current experiments on quantum dots.

The exact solution of the $s$-$d$ model contains a trivial charge sector and
a spin sector with eigenstates determined in the thermodynamic limit
from solutions of linear integral equations.~\cite{alf}
Just as in Fermi liquid theory, a state is labeled by occupation numbers 
$n(i) =0, 1$ on the ``rapidities'' $\lambda_i$ (here the
ordered rapidities $\lambda_i<\lambda_{i+1}$
play the role of momenta in the Fermi liquid theory).
However one complication from 1D interaction is that the position of
rapidities depends on the  occupation numbers $n(i)$.
In the thermodynamic limit, we can introduce the density
of rapidities $D_r(\lambda)=\Delta i/(\lambda_{i+\Delta i} - \lambda_i)$ 
to describe their distribution.
Now a state can be labeled in terms of a
function $\sigma(\lambda)= n[i(\lambda)]D_r(\lambda)$ on the real line
(where $i(\lambda)$ is the integer such that $\lambda_i$ is closest to 
$\lambda$).
Thus $\sigma(\lambda)$ can be interpreted as the density
of occupied rapidities $\lambda$ in the state. The total energy
and magnetization are expressed as weighted integrals of $\sigma$.
All rapidities are filled in the ground state, {\it i.e.} $n(i) =1$
or $\sigma(\lambda)= D_r(\lambda)$. 

The Bethe-ansatz solution is universal in the
limit bandwidth $D \rightarrow \infty$, coupling
$c = 2J/(1-3J^2/4) \rightarrow 0$
with $T_0 = D \exp(-\pi / c)$ finite.~\cite{alf}  The Kondo
temperature $T_0$ is defined as the halfwidth of the DOS peak at
$T=0$, and is related to Wilson's $T_K$ by $T_0 = T_k / w$,
$w = \exp(c+1/4)/\pi^{3/2} = 0.41071\ldots,$ $c$ Euler's constant.

At zero magnetic field, the energy to add a ``hole'' 
at $\lambda_h$ ({\it i.e.}, to set $n[i(\lambda_h)]=0$)
is $\tan^{-1}(\exp(\pi \lambda_h / c))$, and such a hole
has spin $\frac{1}{2}$.
In a magnetic field, 
filling the lowest rapidities with holes lowers energy, {\it i.e.}
$n(i)=0$ for
rapidities $\lambda_i$ in the interval $(-\infty,B)$, with $\exp(\pi B / c) =
\sqrt{e / 2 \pi} (g \mu_B H / T_0)$.
With $B > -\infty$, there are two types of spinons:
hole-like excitations at $\lambda > B$ with $\Delta S = \frac{1}{2}$
and particle-like excitations at $\lambda < B$ (where $n[i(\lambda)]$
is changed from 0 to 1) with $\Delta S = - \frac{1}{2}$.

The density of states for spinon excitations is
$D_s(\epsilon) = D_r(\lambda) (dE / d\lambda)^{-1}$, evaluated at
the value of $\lambda$ with $E(\lambda)=\epsilon$.
The density of rapidities $D_r$ for the ground state (note that $D_r$
depends on the occupation numbers $n(i)$) was found for nonzero $H$ in the
calculation of the impurity magnetization,
so all we need is the energy of the spinon excitation
at rapidity $\lambda$.  The DOS $D_s(\epsilon)$ contains both the
impurity contribution
and a (constant) conduction electron part, ignored henceforth.
The spinon consists of a single $\delta$-function change in the density
$\sigma$ (due to the change in the occupation numbers $n(i)$) plus
the (regular) backflow (due to the change in $D_r$ caused by the
change in $n(i)$).
The backflow $\sigma^\prime(x)$ for a $\delta$-function introduced at
$\lambda_{\rm h} > B$ satisfies
\begin{equation}
\sigma^\prime(\lambda) + \int_B^\infty K(\lambda-\lambda_2)
\sigma^\prime(\lambda_2)\,d\lambda_2 = K(\lambda-\lambda_{\rm h})
\label{backflow}
\end{equation}
with $K(x) = c / \pi (c^2 + x^2)$.
With $B = -\infty$ this becomes, in Fourier space,
${\tilde \sigma}^\prime(p) = 1/(1 + e^{c|p|})$,
so the total density change is
$\delta \tilde \sigma(p) = {\tilde \sigma}^\prime(p) - 1
= -1 / (1 + \exp(-c |p|))$.~\cite{alf}
With a magnetic field, $B > -\infty$ and (\ref{backflow})
can be solved by the Wiener-Hopf method.  The positive backflow
$\sigma^\prime(\lambda)$ now includes both particles ($\lambda > B$) and
holes ($\lambda < B$).  The energy and magnetization of the spinon can be
calculated equivalently by summing over particle or hole states.  The
original hole is dressed by other holes, increasing its interaction
energy but also its magnetization.  The total change in energy as
$\lambda_{\rm h} \rightarrow B^+$ is exactly sufficient
to render the excitation gapless.

The backflow for a hole removed at $\lambda_{\rm e} < B$ satisfies
\begin{equation}
\sigma^\prime(\lambda) + \int_B^\infty K(\lambda-\lambda_2)
\sigma^\prime(\lambda_2)\,d\lambda_2 = - K(\lambda - \lambda_{\rm e}).
\end{equation}
The resulting
``particle-like'' excitation is gapless as $\lambda_{\rm e} \rightarrow B^-$
and connects smoothly to the hole excitation.  For $\lambda_{\rm e} \ll B$,
$\sigma^\prime(\lambda) = -K(\lambda - \lambda_{\rm e})$  and the total number
of holes removed in the backflow is 1.

The Wiener-Hopf technique gives explicit forms for the Fourier transforms
of $\phi_\pm(\lambda) = \sigma(\lambda + B) \theta(\lambda + B).$
The results involve the kernel factorization
\begin{eqnarray}
K_+(x) = (K_-(-x))^{-1} = {(2 \pi)^{1/2} \over \Gamma(1/2 + ix)} \times\cr
\exp[-ix (1 + {i \pi \over 2}
- \log(-x + i \delta))].
\end{eqnarray}
For a hole excitation ($\lambda_{\rm h} > B$):
\begin{equation}
{\tilde \phi_-}(p) = 
{\rm \bf \, P} \int_0^\infty dt\,
{\tan(ct/2) K_+(-ict / 2 \pi) e^{-(\lambda_{\rm h} - B)t} \over
 2 \pi K_-(cp / 2 \pi) (t - ip)}.
\end{equation}
For a particle-like excitation ($\lambda_e < B$):
\begin{equation}
{\tilde \phi_-}(0) = 1 -
\int_0^\infty{dt\,{\sin(ct) \over \pi K_-(0) t}
e^{- (B - \lambda_e) t} K_-(ict/2 \pi)},
\end{equation}
\begin{eqnarray}
{\tilde \phi_-}(i \pi / c) &=& - 
\int_0^\infty{dt\,{\sin(ct)
K_-(ict / 2\pi) e^{-(B - \lambda_e) t} \over \pi K_-(i/2)(t - \pi/c) }} \cr
&\hphantom{=}&-e^{\pi (B -\lambda_e) / c}.
\end{eqnarray}
The total energy for either type of excitation is given by
\begin{eqnarray}
E(\lambda) &=& T_0 e^{\pi B / c} (e^{\pi \lambda / c} +
{\tilde \phi_-}(i \pi / c)) \cr
&& - g \mu H (1 + {\tilde \phi_-}(0)) / 2 \cr
&=& g \mu H (\sqrt{e \over 2 \pi} (e^{\pi \lambda / c} +
{\tilde \phi_-}({i \pi \over c})) -
{1 + {\tilde \phi_-}(0) \over 2}).
\label{energy}
\end{eqnarray}

The ground-state density of rapidities $\sigma_{\rm gs}(\lambda)$ also
satisfies an equation of Wiener-Hopf type with source term
$2 c / \pi (4 \lambda^2 + c^2)$.
The solution for $\sigma_{\rm gs}$ is always peaked near $\lambda = 0$,
with $\sigma_{\rm gs}(\lambda) = 1 / 2 c \cosh(\pi
\lambda / c)$ for $B/c \ll 0$
and $\sigma_{\rm gs}(\lambda) = 2 c / \pi (4 \lambda^2 + c^2)$ for $B/c \gg 0$.
The explicit solution~\cite{andrei,wiegmann} combined with
(\ref{energy}) gives the single-spinon density of states.  The full
(thermodynamic) density of states up to energy $g \mu H$ is a
sum of convolutions of the single-spinon result.  The single-spinon density
of states at zero energy gives the exact spin susceptibility and specific
heat, whose ratio (the Wilson ratio) does not change with magnetic field.~\cite{wiegfink}

\begin{figure}
\epsfxsize=3.0truein
\vbox{\centerline{\epsffile{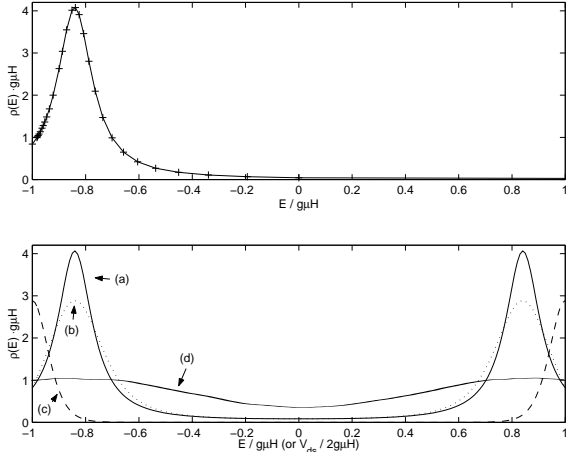}}
\caption{The spin-up (top) and total ((a) in bottom) spectral function estimate
for $g \mu H / T_0 = 48$ ($B / c = 1.1$).  For comparison to the experimental
curve (d) of Goldhaber-Gordon et al.~[16] in a field of 7.5 T from
$V_{ds} = -200 \mu V$ to $200 \mu V$ with $g \mu H / T_0 \approx 50$,
(b) is (a) broadened by a temperature $T = 90$ mK~[16]
and (c) is the EOM result broadened by the same amount.}}
\end{figure}

The spectral function estimate
for spin-up electrons follows from the single-spinon
density of states by taking $\rho_\uparrow(\omega) = \frac{1}{2}
D_\sigma(|\omega|)$ with
$\sigma = +\frac{1}{2}$ for $\omega > 0$ and $\sigma = -\frac{1}{2}$ for
$\omega < 0$, where $\sigma$ is the spin relative to the ground state.
States of spin $+\frac{1}{2}$ above the ground state
can be obtained equivalently by
adding a spin-up electron or by subtracting a spin-down
electron, so the Kondo part of the
spectral function is always symmetric under the combined
transformation $H \rightarrow -H, E \rightarrow -E$.

The shape of the spectral function is determined by the ratio $g \mu H
/ T_0$.  Fig. 1 shows the spin-up and total spectral functions for $g
\mu H / T_0 = 48$.  The peak in the single-spin spectral function has
width $2 T_0$ at zero field and shifts away from the Fermi level and
broadens with increasing field (Fig. 2).  However, the width of the
peak decreases as a fraction of $g \mu H$.  The shift of the peak is
always greater than the noninteracting level value $g \mu H / 2$,
starting at about two-thirds $g \mu H$ for small $g \mu H/T_0$ and
rising slowly to $g \mu H$ as $g \mu H / T_0 \rightarrow \infty.$ The
effect of temperature on the equilibrium spectral function will be
slight until $T \geq T_0$; the effect of high temperature will be to
blur of the Kondo correlations (and hence broaden and shift of the
peaks toward the EOM result), but this is a smaller effect
experimentally than the nonequilibrium broadening.  The behavior of the
finite-$T$ magnetization suggests that the temperature needs to be a few
times $T_0$ for a strong effect.

\begin{figure}
\epsfxsize=3.0truein
\vbox{\centerline{\epsffile{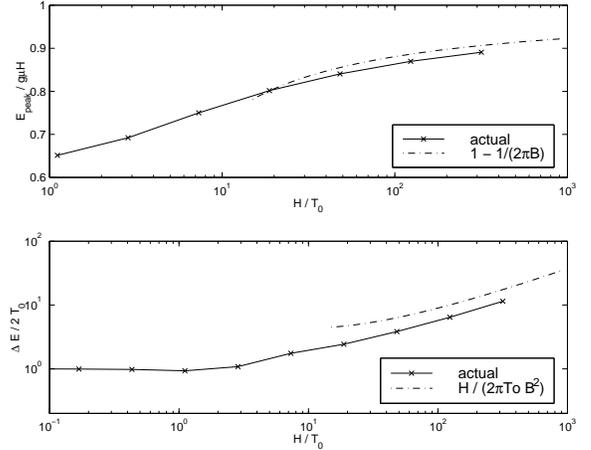}}
\caption{The peak shift (top) in units of $g \mu H$
and broadening (bottom) in units of $2 T_0$, for various values of the ratio
$g \mu H / T_0.$  The dotted lines are limiting forms (12) and (13).}}
\end{figure}

The spectral function estimate can be compared to existing
measurements of the $H>0$ differential conductance through a Kondo
impurity.~\cite{ralph,goldhaber,cronenwett} Within the NCA at $H=0$,
the spectral function for nonzero bias $V_{ds}$ is essentially two
broadened copies of the $V_{ds}=0$ spectral function, one shifted up
by $V_{ds}/2$ and one down by $V_{ds}/2$.~\cite{mwl} Note that
$V_{ds}$ is the applied (drain-source) voltage across the dot, rather
than the gate potential.  The thermal broadening ($T = 90$ mK) and
additional nonequilibrium broadening in \cite{goldhaber} mean that the
observed peak is wider than the zero-temperature prediction, but the
inward shift of the peak center should still be visible, and in
\cite{ralph} and \cite{goldhaber} a smaller splitting than the EOM
result of \cite{mwl} was observed.  Fig. 1 shows the agreement of our
calculated peak location with that measured in \cite{goldhaber} for
$|g| = 0.36$, the value suggested by electron spin-resonance data on
2DEGs~\cite{dobers} for the 7.5 Tesla field in that experiment.  The
value $|g| = 0.30$ required for agreement with the EOM result is less
likely.  The experimental peaks in Fig. 1 show some nonequilbrium
broadening~\cite{mwl}, but the parameters needed to estimate this
broadening were not measured; convolving the theoretical curve with a
Lorentzian of variable width gives a very good fit to experiment.


Because $g$ in heterostructures is difficult to measure
directly, a more convincing experimental demonstration of the splitting
could be obtained by varying $g \mu H / T_0$.  In experiment
\cite{cronenwett} the nonequilibrium broadening is relatively large and
may blur the correlations enough that the EOM result becomes
applicable.

An asymptotic formula for the peak shape in the limit of large magnetic field
can be derived from the large-$B$ limit of (\ref{energy}) and
$\sigma_{\rm gs}(\lambda) \approx 2 c / \pi (4 \lambda^2 + c^2):$
\begin{equation}
{E(\lambda) \over g \mu H} = 1 - {1 \over 2 \pi x}
+ O(x^{-2} \log(x)),\quad x = B - \lambda.
\end{equation}
The result for the peak location in this limit is
\begin{equation}
\label{peakloc}
{E_{\rm max} \over g \mu H} \approx 1 - {1 \over 2 \pi B} =
1 - {1 \over 2 \log(g \mu H \sqrt{e} / T_0 \sqrt{2 \pi})}.
\end{equation}
In the large-$H$ limit the split between spin-up and spin-down peaks
in the spectral function is $2 g \mu H$ since spin-up and spin-down
peaks each shift by $g \mu H$.  The relative correction in (\ref{peakloc})
is half that in the impurity magnetization $M \approx (g \mu H / 2) [1 -
1 / \log(g \mu H \sqrt{e} / T_0 \sqrt{2 \pi})].$

The peak width (FWHM) in the limit $H \gg T_0$ is
\begin{equation}
\label{peakwid}
\Delta E \approx {g \mu H \over 2 \pi B^2} =
{g \mu H \over 4 \log^2(g \mu H \sqrt{e} / T_0 \sqrt{2 \pi})}.
\end{equation}
The narrowness of the peak compared to $g \mu H$
in the large-$H$ limit is consistent with the EOM results.~\cite{mwl}

There are two subtleties in the calculation worth mentioning.  The
above expression for the magnetization gives a value continuously
varying from $\Delta M = 1/2$ for hole-like excitations with $\lambda
\gg B$ to $\Delta M = 1$ for particle-like excitations with $\lambda
\ll B$.  There are not continuous-spin excitations; rather the
physical excitation combines a dressed spinon with an infinitesimal
shift of the chemical potential for spinons $B$ to give an integer or
half-integer total spin.  The same phenomenon of apparently continuous
quantum numbers was seen in the charge sector of the asymmetric
Anderson model.~\cite{kawakami} At nonzero $H$ and $T=0$ there is a
small jump in the single-spinon density of states at $E = g \mu H$.
This discontinuity is balanced by the appearance of new types of
excitations, so that the total (thermodynamic) density
of states is continuous but for a $\delta$-function at $E = g \mu
H$ from a state in the same $SU(2)$ multiplet as the
ground state.

We now suggest a reason why large-$N$ techniques like the NCA
give a spurious peak at the Fermi level for finite $H$.
Recall that the impurity orbitals
transform under $SU(N)$ rather than in
a high-$S$ representation of $SU(2)$. The $N$ degenerate
orbitals split in a generalized magnetic field as follows: $N/2$ shift
up by $g \mu H / 2$ and $N/2$ shift down by the same
amount.~\cite{represent}
The $N \rightarrow \infty$ magnetization curves for this
and other splittings are derived in~\cite{withoff}.
The Kondo peak near the origin then arises from
transitions between states at the same energy, while the split peaks
arise from transitions between states at different energy.
Counting the weight of each peak in this argument gives the
correct value of the spin susceptibility Wilson ratio,
$N / (N-1)$.~\cite{noz}
The $N = 2$ problem is qualitatively different than the $N \geq 4$
problem since there are no degenerate orbitals after the magnetic field
is applied, so it is not surprising that the analytic continuation down
to $N = 2$ gives a false peak at the origin, which is a real peak for
$N \geq 4.$ 
The problem of the conductance through a quantum dot in the Kondo regime
is a fine example of the subtle properties of nonequlibrium
transport in strongly correlated systems.

J. E. M. acknowledges support from the Hertz Foundation and
X.-G. W. from NSF grant DMR98-08941.  {\bf Note added in proof:} A
preprint by T. Costi (cond-mat/0004037) finds comparable results for
 $H \leq T_0$ from an NRG calculation of the spectral
function.


\begin{references}
\bibitem{represent}{Which representation at general $N$ is selected by the
$N=2$ calculation is in general difficult to determine, but for $N = 2^n$ the
appropriate representation is the spinor representation (outer products of
Pauli matrices) which splits as described.}
\end{references}
\end{document}